%% file: main.tex
\documentclass[10pt, conference, a4paper]{IEEEtran}
\IEEEoverridecommandlockouts
\pdfoutput=1 
\usepackage[left=1.65cm,right=1.65cm,top=1.9cm]{geometry}

\usepackage{mathtools}
\usepackage{comment}

\usepackage[utf8]{inputenc}
\usepackage{graphicx}
\usepackage{adjustbox}
\usepackage{listings}
\usepackage{enumitem}
\usepackage{multirow}
\usepackage{subfigure}
\usepackage{amsmath, amsthm, amssymb}
\usepackage{tikz}
\usepackage{framed}

\usepackage{url}
\usepackage{syntax}
\usepackage{balance}
\usepackage[most]{tcolorbox}
\usepackage{newfloat}
\usetikzlibrary{shadows}
\usepackage{tikz-dependency}

\definecolor{meublue}{rgb}{0.37, 0.35,0.97}
\definecolor{meugreen}{rgb}{0.25, 0.75, 0.25}
\definecolor{codegray}{rgb}{0.95, 0.95, 0.95}

\newcommand*\circled[1]{\tikz[baseline=(char.base)]{        
 \node[shape=circle,fill,inner sep=0.8pt] (char) {\textcolor{white}{#1}};}}

\usepackage{tcolorbox}

\usepackage{amssymb}
\usepackage{pifont}

\tikzset{
every label/.style={xshift=-4ex, text width=6ex, align=right, 
                             inner sep=1pt, font=\footnotesize, text=red},
    my node style/.style={
        font=\small,
        rectangle,
        minimum size=3mm,
        draw=black!75,
        very thin,
        fill=white!4,
        align=center,
    }
}

\newcommand{\sysname}{Araucaria}

\AtBeginEnvironment{grammar}{\small}

\DeclareFloatingEnvironment[
  fileext   = logr,
  listname  = {List of Grammars},
  name      = Grammar,
  placement = htp
]{Grammar}

\lstdefinelanguage{Resist}{
language=c,
  keywords=[1]{action, if},            
  keywordstyle=[1]\bfseries,       
  keywords=[2]{@, Create, Update, Synchronous, intent, def, operation, func, meta}, 
  keywordstyle=[2]\bfseries,       
keywords=[3]{functionality, consistency, availability, tolerates, failures, priority, Input, nodes, require, igmd, iter0, iter1, tmin, min}, 
  keywordstyle=[3]\color{violet},       
}

\def\BibTeX{{\rm B\kern-.05em{\sc i\kern-.025em b}\kern-.08em
    T\kern-.1667em\lower.7ex\hbox{E}\kern-.125emX}}
\begin{document}

\title{\sysname{}: Simplifying INC Fault Tolerance \\ with High-Level Intents}

\author{
\IEEEauthorblockN{
	Ricardo Parizotto\IEEEauthorrefmark{2},
	Israat Haque\IEEEauthorrefmark{1},
  		Alberto Schaeffer-Filho\IEEEauthorrefmark{2}},

\IEEEauthorblockA{
	\IEEEauthorrefmark{1}\textit{Dalhousie University},		\IEEEauthorrefmark{2}\textit{UFRGS}
}}

\maketitle

\begin{abstract}
Network programmability allows modification of \textit{fine-grain} data plane functionality. The performance benefits of data plane programmability have motivated many researchers to offload computation that previously operated only on servers to the network, creating the notion of \textit{in-network computing} (INC). Because failures can occur in the data plane, fault tolerance mechanisms are essential for INC. However, INC operators and developers must manually set fault tolerance requirements using domain knowledge to change the source code. These manually set requirements may take time and lead to errors in case of misconfiguration. In this work, we present \sysname{}, a system that aims to simplify the definition and implementation of fault tolerance requirements for INC. The system allows requirements specification using an intent language, which enables the expression of consistency and availability requirements in a constrained natural language. A refinement process translates the intent and incorporates the essential building blocks and configurations into the INC code. We present a prototype of \sysname{} and analyze the end-to-end system behavior. Experiments demonstrate that the refinement scales to multiple intents and that the system provides fault tolerance with negligible overhead in failure scenarios. 
\end{abstract}

\input{introduction}
\input{background}
\input{requirements}

\input{design}

\input{evaluation}

\input{related\_work}
\input{conclusions}

\balance

\bibliographystyle{plain} 
\bibliography{sample-base}

\end{document}

%% file: introduction.tex
\section{Introduction}

Network programmability has changed how we manage and operate computer networks, providing agility for deploying new network functions in the network. The P4 language \cite{Bosshart:2014} has paved the way towards programmability, allowing fine-grain data plane functionality development. This programmability has motivated the notion of \textit{in-network computing} (INC) \cite{10.1145/3152434.3152461}. INC advocates for offloading functionality traditionally running on servers to programmable network devices. This offloading has advantages, e.g., reducing latency and improving bandwidth by intercepting and processing network packets at the switch; thus avoiding the need to forward them to servers. Examples of INCs in the literature include functionalities such as aggregation \cite{265065}, load balancing \cite{barbette2021cheetah, tajbakhsh2022accelerator}, concurrency control \cite{li2017eris}, computer vision \cite{glebke2019towards, 9763800} and IoT security \cite{kuzniar2022iot}.

However, these advantages come at the cost of data plane configuration complexity, due to the necessity of writing low-level P4-based  operations (e.g., table entries, action parameters, and register values). Configuring such functions in the data plane is tedious and requires substantial training. Moreover, data plane programmability needs to deal with failures in forwarding devices and in the applications running in them, which adds an additional layer of complexity. For example, existing fault tolerant systems \cite{kim2021redplane, zeno2022swish} employing replication techniques on data plane devices introduce additional complexities. In particular, they require configuring an INC and its replicas while offering consistency among these replicas to offer fault tolerance.

One way to mitigate the above complexities is to develop a DevOps-friendly automated policy- or intent-based data plane configuration. This approach could enable INC operators to express fault tolerance requirements at a higher abstraction level \cite{pang2020survey}. Subsequently, the underlying software would automatically translate these specifications into detailed low-level data plane code and configurations. Although early research efforts in policy management facilitated some configuration capabilities, particularly in the domain of security and quality of service (QoS) \cite{damianou2001ponder}, they can not enable the specification of policies for programmable data planes. Recent developments in \textit{intent-based networking }(IBN) allow the deployment of high-level intents directly into more fine-grain network configurations, such as OpenFlow \textit{match+action} rules, middleboxes \cite{jacobs-2018, 8725754, 10.1145/3281411.3281431, alalmaei2020sdn}, or P4 \cite{angi2022nlp4,8526852,riftadi2019p4i} -  an in-depth list is presented in Table~\ref{tab:Taxonomy}. However, existing works in this domain do not support  fault tolerance requirements and abstractions. Consequently, adding fault tolerance to INC still requires handling low-level switch code without a clear and organized methodology.

\begin{table}[h!]
\centering
\caption{Intent frameworks in the literature.}
\label{tab:Taxonomy}
\resizebox{.47\textwidth}{!}{%
\begin{tabular}{l|c|c}
\multicolumn{1}{c|}{\textbf{Examples}}      & \textbf{Purpose}      & \textbf{Target} \\ \hline \hline
\begin{tabular}[c]{@{}l@{}}Nile \cite{jacobs-2018}, InsPire \cite{scheid2017inspire}, \\ PGA \cite{Prakash:2015:PUG:2785956.2787506},\\  Arkham \cite{machado2017arkham}\end{tabular} &
  QoS, Control Access &
  SDN, NFV \\ \hline
Janus \cite{10.1145/3143361.3143380} &
  \begin{tabular}[c]{@{}c@{}}Bandwidth QoS,\\ Temporal Policies\end{tabular} &
  SDN, NFV \\ \hline
JingJing \cite{10.1145/3341302.3342088}     & ACL rules             & WAN             \\ \hline
Gherkin \cite{8725754}                      & Firewall, SFC         & SDN             \\ \hline
P4-iO \cite{riftadi2019p4i}, \cite{8526852} & Routing, HH Detection & P4             
\end{tabular}%
}
\end{table}



 
In this work, we propose \sysname{}\footnote{Araucaria is named after the \textit{Araucaria angustifolia}, which are large and resilient trees that we can be found in the south of Brazil. The Araucaria is a symbol of \textit{resistance} in the fight for biodiversity conservation.}, a system that utilizes the expressiveness benefits of IBN for providing fault tolerance for INC. \sysname{} enables operators to specify fault-tolerance requirements using a high-level and declarative language. We introduce a \textit{refinement process} that translates the operator intents into data plane code and configurations to preserve the requirements for INC. This refinement process comprises three main steps: translation, instrumentation, and configuration. Initially, \sysname{} \textit{translates} the intent into an intermediary representation, identifying the essential building blocks to achieve the desired level of robustness. Next, \sysname{} \textit{instruments} the INC source code with a set of constructs  (e.g., the parser, definition of headers, and control flow) to enforce a replication protocol that satisfies the high-level fault tolerance intent. Finally, the system \textit{configures} the data plane according to the translated intent and network topology information. The configuration includes several rules, such as multicast groups, ports, IP addresses, and register entries. These rules ensure replicas preserve the consistency model necessary and recover appropriately from failures. The instrumented code can then be automatically deployed and configured according to the intent priority. 

To validate our approach, we have implemented a prototype of \sysname{} both for the behavior model of P4 programs (BMv2) and for a Tofino switch ASIC (Edgecore Wedge 100BF-32X). Additionally, we conducted experiments to assess the scalability of our refinement process. As a practical case study, we evaluate \sysname{} by analyzing the system behavior when injecting failures, and show that \sysname{} can rapidly allow INCs to recover from failures. 



\textbf{Contributions.} Overall this paper makes the following contributions:
\begin{itemize}
    \item Identifies a set of requirements and abstractions to provide fault tolerance for in-network computing.
    \item Proposes a refinement technique to systematically instrument the INC source code with fault tolerance protocol building blocks, ensuring consistency guarantees. 
    \item Evaluates a use case with an existing INC and analyses results in terms of feasibility and scalability. Results show that \sysname{} can recover from failures in less than $0.2s$, while also keeping INC consistency. 
\end{itemize}


%% file: background.tex
\section{Background}

In this section we discuss the essential background for intent-based networking and programmable data planes. 



\subsection{Intent-based Networking (IBN)} Intent-based networking simplifies network management through five essential components: profiling, translation, resolution, activation, and assurance \cite{rfc9315}. Users specify their abstract intentions in the profiling phase using high-level representations,  such as natural language. Subsequently, the translation component refines these into policies and further into concrete configuration commands \cite{elkhatib2017charting}. The resolution component solves conflicts, ensuring a conflict-free activation of the intent configuration onto network devices. Despite the reliance on translation and conflict resolution for deploying intents, the dynamic nature of the network may lead to configurations no longer satisfying the initial intent. To address this challenge, an intent assurance component employs mechanisms to identify the discrepancies and refine new configurations to deploy in the network \cite{leivadeas2022survey}.

\begin{figure}[ht] 
 \vspace{7px}
\centering\includegraphics[width=0.33\textwidth]{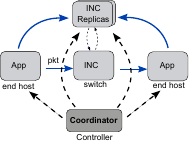}
    \caption{Fault Tolerant INC system model.}
    \label{fig:faultback}
\end{figure}

\subsection{In-Network Computing} 

In-network computing (INC) has been used to characterize systems with functionality offloaded to the data plane of networking devices \cite{michel2021programmable}. INC relies on programmable switches or SmartNICs that are already deployed on the traffic path~\cite{25019} to potentially reduce the need for additional specialized equipment, such as accelerators or middleboxes. INC is mainly motivated by the performance benefits it can provide. Those benefits include reducing latency and bandwidth usage, increasing throughput, and improving energy consumption~\cite{25019}. \looseness=-1

However, these advantages come with a cost: once we offload functionality to the data plane of networking devices, failures in these devices may affect the correctness of the system. Fault tolerance is thus essential but adds additional complexity to configure the INC. 

%% file: requirements.tex
\section{INC Fault Tolerance}

To better understand fault tolerance requirements, we illustrate a hypothetical INC fault tolerance model in Figure \ref{fig:faultback}. In particular, a switch running an INC intercepts packets from applications running on end hosts. The switch synchronizes the INC state with a pre-configured set of replicas and is able to recover from crashes. The fault tolerance of the INC is measured as reliability (e.g, by means of consistency) and availability of the INC when it is necessary. 

\textbf{Replication for availability.} Redundancy is a common approach to ensure the system is available even in case of failures. Multiple INC replicas can keep replicated state, where replication can take different forms: it can be either \textit{synchronous} in a way that resembles the execution of a sequential processor \cite{herlihy1990linearizability}, or \textit{asynchronous}, allowing temporary inconsistency between replicas \cite{kim2021redplane}. In case of a switch crash, a coordinator can identify the failure and orchestrate the recovery by collaborating with replicas and end-hosts. The recovery process can also take different forms, from simply \textit{rerouting} new application requests to the replicas, to a more complex process of \textit{replaying} packets from the applications to restore the state of inconsistent replicas.

\begin{figure*}[ht]
 \vspace{7px}
    \centering
    \includegraphics[width=0.75\textwidth]{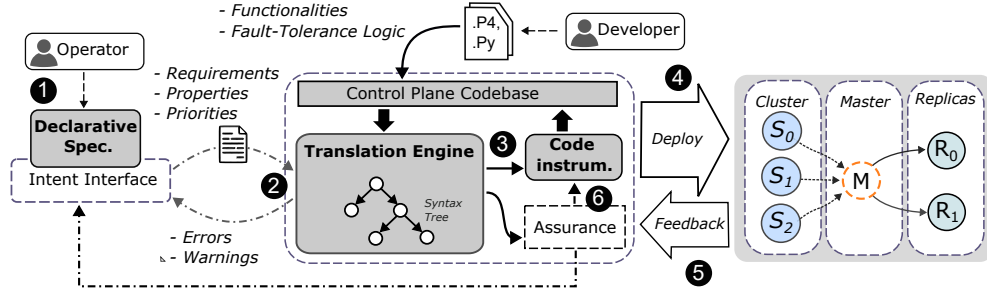}
    \caption{The high-level architecture of Araucaria.}
    \label{fig:Araucaria}
\end{figure*}

\textbf{Consistency notions.} Replicas can process packets in an \textit{ordered} or \textit{unordered} manner, impacting the application's consistency. Different types of applications may be correct under different consistency models, such as strong or eventual. On the one hand, \textit{strong consistency} ensures replicas process all the requests in the same total order as the primary INC, while \textit{eventual consistency} does not require strict ordering. In addition, specific applications can achieve a strong notion of consistency without ensuring strict ordering \cite{zeno2022swish}. These applications follow a system model called \textit{strong eventual consistency} (SEC), in which requests are processed as they arrive, and conflicts are solved automatically using a merge function. The data replicated from these applications are called \textit{conflict-free replicated data types} (CRDT) and often apply to independent or monotonically increasing data \cite{shapiro2011conflict}. Examples are operations in different keys of a key-value store or adding values to a distributed counter, e.g., counting likes in social media posts. 





%% file: design.tex
\section{\sysname{} Design}
\label{intent:challenges}

In this section, we present the design of \sysname{}, a system that relies on intents to enhance the fault tolerance of INCs. The design of \sysname{} is based on the following insights.  \looseness=-1

\textbf{Fault tolerance specification.} To simplify INC fault-tolerance, the specification of intents should abstract the implementation details. Intents must also be expressive enough to fulfill different fault tolerance requirements. To solve these issues, \sysname{} defines a constrained natural language with primitive fault tolerance constructs, simplifying the specification of fault tolerance requirements.

\textbf{Systematic instrumentation.} Instrumenting fault tolerance into INC is difficult because of the limited composability of the P4 language -- i.e., simply importing the fault tolerance functionality is impossible. To solve this challenge, \sysname{} provides a refinement methodology that deduces the rules to provide fault tolerance from the input intent. The system also defines an instrumentation strategy capable of systematically instantiating fault tolerance building blocks into the INC source code.  

\subsection{Overview and workflow}
\label{intent:methodology}

Figure \ref{fig:Araucaria} illustrates the overview and workflow of Araucaria. Initially, the operator defines INC fault tolerance requirements (e.g., in terms of \textit{consistency notion} and \textit{number of replicas}) in a declarative manner (\circled{1}). The specification, made using a high-level language, goes through a translation process that analyzes the intent structure and semantics (\circled{2}). If the translation occurs without errors, \sysname{} generates an intermediary representation, identifying pre-defined building blocks that implement the fault tolerance logic. For example, these building blocks include code fragments for enforcing \textit{failure detection} or \textit{packet replaying} mechanisms. \sysname{} then instruments the INC code by merging parsers and control flows from the fault tolerance building blocks into a single data plane program (\circled{3}). This data plane program is instantiated into multiple switch replicas based on the required availability (\circled{4}), and an assurance module is instantiated in the control plane to coordinate fault recovery. Dynamically, \sysname{} replicates the INC state across devices and provides periodic feedback to the assurance module about the status of the replicas (\circled{5}). In the event of an INC failure, the assurance module refines the data plane configurations (\circled{6}), forwarding application packets to a different replica INC.  \looseness=-1

\subsection{Declarative intent specification} 
\label{sec:intentspec}

An \textit{intent} is an abstract declaration of what an application or user desires from the network \cite{tsuzaki2017reactive}. In \sysname{}, each intent is associated with a predicate, including functionality, requirements, and priorities. These predicates state a \textit{property} of an intent. Inspired by \cite{elkhatib2017charting, jacobs-2018}, we formulate a constrained natural language to specify fault tolerance intents, where intents are structured as a tuple of primitive elements \textit{$\langle$operations, functionalities, requirements$\rangle$}. Grammar \ref{gra:sample} presents the language specification. 

\setlength{\grammarparsep}{0.04cm}
\begin{Grammar}[h!]
 \vspace{5px}
\begin{grammar}

<intent>     ::= <op> \texttt{intent\_name}  `{' <pred> `}'

<op> ::= `Create' | `Delete' | `Update' | `Read'

<pred> ::= { <req> `,' <func>  }

<func> ::= \texttt{functionality} `fname' `[' <input> `]' `,' 

<reqs>    ::= <reqs> ',' <req>  | <req> 

<req>       ::= <avail> | <cons> | <cons> `[' <merge> `]' 

<inputs> ::= <inputs> `,' <input> | <input> | <empty>

<input> ::=  \texttt{name} `:' \texttt{value}

<avail>  ::= \texttt{tolerates} <int> `failures'

<cons>   :: = `strong' | `eventual' 

<merge>  ::= \textbf{max[hdr.value)]} | `add'



\end{grammar}
 \caption{The Araucaria grammar in BNF.} \label{gra:sample}
\end{Grammar}



The language constructs are:
\begin{itemize}
    \item \textbf{\textit{Operations}} define actions (Create, Read, Update, and Delete) being applied to instances of \textit{functionalities}. 
    \item  \textbf{\textit{Functionalities}} identify the specific INC that the intent aims to configure. Functionalities may be instantiated with customized \textit{inputs} that are used during the refinement to identify the necessary INC building blocks for deployment.  \looseness=-1 
    \item \textit{\textbf{Requirements}} is the core element in the \sysname{} intent structure. A requirement aims to provide additional information about the intent: 

    \begin{itemize}
        \item \textit{Availability} lets programmers ensure that specific INCs are available even if $f$ failures occur \cite{cheung2021new}. We assume failures can occur by crashing, but switches do not experience an arbitrary behavior (i.e., no byzantine cases).

        \item \textit{Consistency} allows programmers to specify replica correctness properties. The properties can vary between a strong or weaker notion that does not preserve ordering constraints. In addition, consistency may be followed by an optional merge function that provides ways to reduce conflicts between requests. 
    \end{itemize}

\end{itemize}

Listing \ref{lst:simulation} presents an example of an intent that can be built using the \sysname{} intent language. In this example, an intent called `\texttt{syncIntent}' is created. The synchronization functionality expects an optional parameter representing the number of processes interacting with the network functionality. The intent requires the INC to tolerate two simultaneous failures while preserving strong consistency.
 
\begin{lstlisting}[escapechar=!, belowskip=1 \baselineskip, float=htpb, caption={Intent for synchronization functionality.},captionpos=b, label={lst:simulation} ]
Create intent syncIntent{
    functionality: synchronization [
                      size: 3
                ]
    availability: tolerates two failures,
    consistency: strong,
} 
\end{lstlisting}

We implemented a compiler to translate \sysname{} intents, including a \textit{lexer} (to identify the tokens from the intent) and a \textit{parser} (to analyze the syntactic structure of the intent and generate an abstract syntax tree (AST)). Also, the \textit{semantic analysis} ensures correct input formatting and examines potential conflicts, such as assessing whether the expressed merge function can achieve the desired consistency mode. The output of the intent compilation process is either an error or a valid intent represented at a lower level. This representation contains the fault tolerance functionality decomposed into smaller building blocks, which are discussed next.  \looseness=-1

\subsection{Fault tolerance building blocks}

\begin{figure}[t!]
 \vspace{7px}
    \centering
    \includegraphics[width=0.45\textwidth]{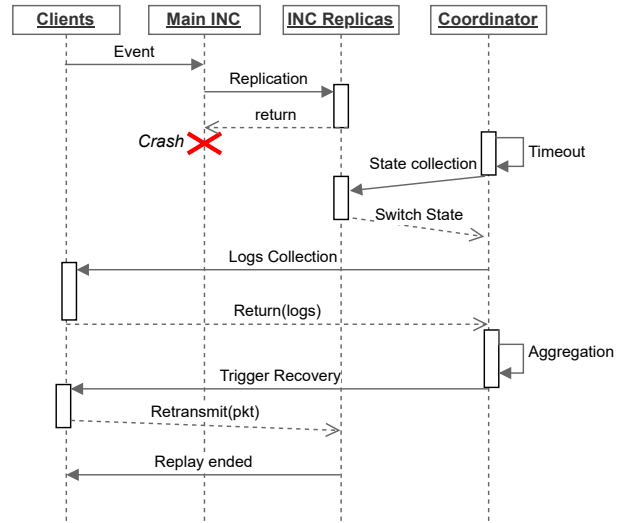}
    \caption{Sequence diagram for \sysname{} Recovery protocol in accordance with Figure \ref{fig:faultback}.}
    \label{fig:protocol}
\end{figure}

\sysname{} defines a fault tolerance protocol for recovering INCs from failures. In this protocol (Figure \ref{fig:protocol}), client traffic is processed by the main INC and replicated to a set of switch replicas. If the main INC crashes, a control plane program (\textit{Coordinator}) identifies the failure using timeouts and collects the necessary state information from the replicas and the clients. The Coordinator can aggregate their information and identify the subset of packets that need to be retransmitted to a replica. The aggregated information may trigger a client replay, which recovers the replicas to a consistent state~\cite{park2019exploiting}. 

The refinement process implemented by \sysname{} merges the source code of pre-existing building blocks that enforce the fault tolerance protocol with the INC source code. Dynamically, the network operator can select one of the recovery strategies instrumented into the INC according to an application's consistency requirement, i.e., strong or weak.

\begin{figure}[h!]
    \centering
    \includegraphics[width=0.45\textwidth]{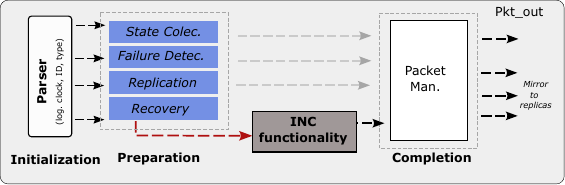}
    \caption{Structure of an INC instrumented with fault tolerance building blocks.}
    \label{fig:structure}
\end{figure}

\textbf{Reusable building blocks.} Figure \ref{fig:structure} provides a comprehensive overview of the underlying structure of a P4 program that has been instrumented with \sysname{} to support fault tolerance employing a set of four standard building blocks:  \textit{Failure Detector}, to identify if the main INC has failed; \textit{Replication}, to synchronize state with other switches; \textit{State Collection}, to determine how up to date a replica state is; and \textit{Recovery}, to handle the recovery ensuring replicas follow a specific consistency notion after a failure. 

These building blocks are implemented as a set of P4 \textit{templates} that are merged with the INC source code. Templates include three blocks: \textit{initialization}, \textit{preparation}, and \textit{completion}.

\begin{itemize}
    \item \textbf{Initialization.} The initialization template includes per-packet variables, such as custom metadata, a new header and struct, and a parser state. The header has information to identify servers and message types (e.g., recovery, collection) and to ensure linearizability through monotonically increasing logical clocks. The parser state can initialize these variables upon the arrival of a packet.  
    
    \item \textbf{Preparation.} This block includes a set of variables to implement the building blocks we mentioned earlier. The Preparation template precedes the INC functionality in the pipeline and consists of code that prepares the packet for INC processing or filtering. Filtering is essential in several cases, such as when the packet being handled is an acknowledgment for replication or a message for failure detection. In these cases, the INC itself should refrain from processing these packets.
    
    \item \textbf{Completion.} This block includes packet management mechanisms capable of applying multicast tables, keeping storage for packet losses, and changing header arguments. This block should be included after the INC functionality to preserve headers and packet metadata for correct INC processing. 
\end{itemize}

\subsection{INC source code instrumentation}

To instrument the \textit{templates} discussed in the previous section into an INC, \sysname{} systematically traverses the INC code and writes \texttt{include} pre-processors strategically to instantiate the building blocks in distinct parts of the INC source code. We require INC variables to follow a specific naming convention to avoid conflicts with variable names used by \sysname{}. This requires that INC variable names do not start with  \sysname{} \textit{reserved words}, thereby establishing a contract between INC developers and DevOps.

\begin{figure}[t!]
 \vspace{4px}
    \centering
    \begin{minipage}{.24\textwidth}
        \centering
        \includegraphics[width=0.98\textwidth]{figures/araucaria\_parser.pdf}
        \caption{Fault tolerance parser.}
        \label{fig:parserResist}
    \end{minipage}%
    \begin{minipage}{0.24\textwidth}
        \centering        \includegraphics[width=0.88\textwidth]{figures/parser\_netgvt.pdf}
        \caption{INC parser.}
        \label{fig:parserGVT}
    \end{minipage} \\
    \begin{minipage}{0.24\textwidth}
    \vspace{0.4cm}
        \centering       \includegraphics[width=0.93\textwidth]{figures/composed\_parser.pdf}
        \caption{Instrumented parser.}
        \label{fig:parserComposed}        
    \end{minipage}
\end{figure}

\textbf{Step \#1: Metadata and header definitions.} The first phase of the instrumentation includes the definitions of \texttt{headers} and \texttt{structs} at the beginning of the INC source code. \sysname{} variables include a new header definition for ensuring linearizability during replication and specific metadata used in the control flow for making per-packet decisions. We start instrumenting the parser after merging the INC headers, \texttt{structs}, and metadata definitions.

\textbf{Step \#2: Parser instrumentation.} Our parser instrumentation leverages a modular design that decouples the \sysname{} parser state from traditional protocols such as Ethernet and IPv4. This decoupling allows us to incrementally include the \sysname{} protocol into the INC parser, avoiding ambiguities. This is achieved in two steps: first, placing the \sysname{} state between the INC header extraction and the transitions, with the INC state working as the `parent' node of the \sysname{} state. Additionally, the \sysname{} state incorporates previous INC state transitions. By ensuring that the extraction of an INC state is consistently followed by the extraction of the \sysname{} header, we effectively mitigate the risk of introducing loops and non-determinism in the parser structure.  \looseness=-1

Figure \ref{fig:parserResist} presents the parser of \sysname{}, ignoring the states for standard protocols. Figure \ref{fig:parserGVT} shows a general INC parser, including the INC state. During the instrumentation, \sysname{} includes the transitions from the INC in a transition of the \texttt{Parse\_Araucaria}. The INC state transitions are also removed, adding a single transition to the \sysname{} state. The resulting parser is presented in Figure \ref{fig:parserComposed}.

\textbf{Step \#3: Control flow composition.} After instrumenting the INC parser, we start instrumenting the control blocks. Control blocks in  P4 can contain several constructs, such as \textit{tables}, \textit{actions}, \textit{registers}, and \textit{apply blocks}. To compose the INC source code with the fault tolerance logic, \sysname{} extends the definition of tables, actions, and registers in the INC code to offer consistency. These include variables for serializing requests between replicas, keeping consistency, and actions to handle packets from the replica and coordinator. Next, \sysname{} proceeds to instrument the source code within the apply block. Our approach includes the entire INC apply block between the \textit{preparation} and \textit{completion} templates. This will allow, for example, to identify unordered packets in the preparation template to avoid processing them in the INC code. Another example is to ensure that multicasts in the completion template do not process packets within the same switch.  \looseness=-1

\subsection{Configuration}

Beyond instrumenting the source code of the INC for a specific intent, \sysname{} generates the configuration to the network devices. The configuration \sysname{} creates is responsible for different tasks: (1) setting up the switch ports for replicating packets; (2) configuring the switches and servers to operate accordingly to a specific consistency model; and (3) setting up the communication with all the servers running applications using the INC. 

\begin{figure*}[ht!]
    \begin{minipage}{.33\textwidth}
    \centering
    \includegraphics[width=0.98\textwidth]{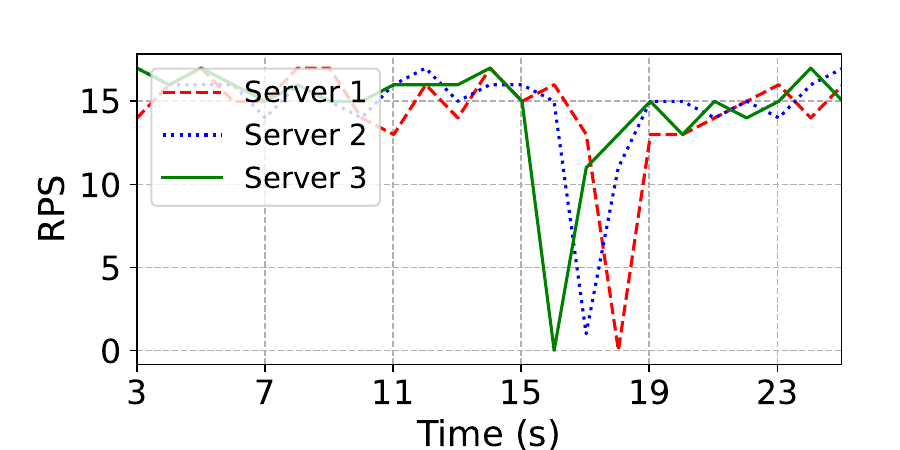}
    \vspace{-0.3cm}
    \caption{Analysing the behavior after failure.}
    \vspace{-0.3cm}
    \label{fig:emulation}
    \end{minipage}%
    \begin{minipage}{0.33\textwidth}
    \centering
\includegraphics[width=0.98\textwidth]{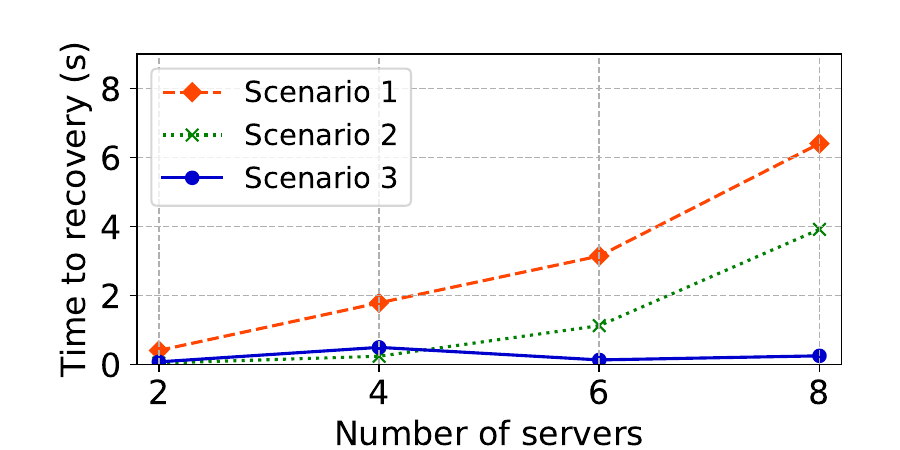}
\vspace{-0.3cm}
    \caption{Recovery on different amounts of servers.}
    \vspace{-0.3cm}
    \label{fig:Crashrecovery}        
    \end{minipage}%
    \begin{minipage}{0.33\textwidth}
    \centering
\includegraphics[width=0.98\textwidth]{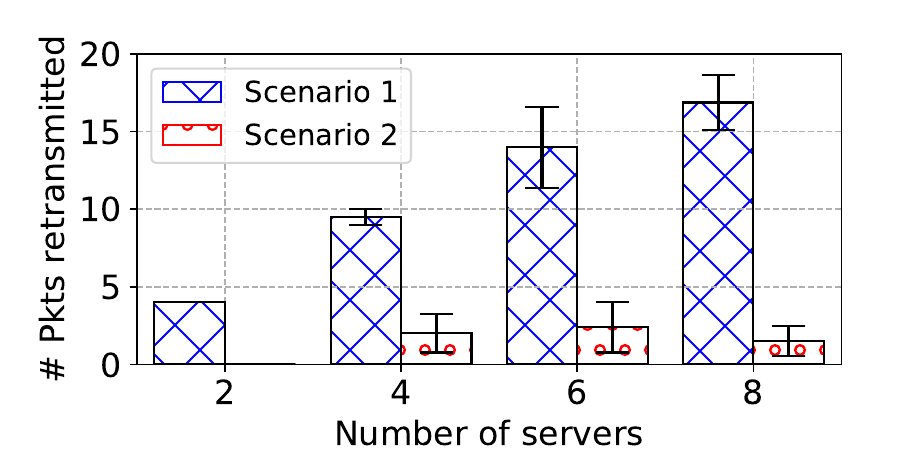}
\vspace{-0.3cm}
    \caption{Number of packets to replay.}
    \vspace{-0.3cm}
    \label{fig:replay}   
    \end{minipage}%
\end{figure*}

\begin{itemize}
    \item \textbf{Replication for availability.} The availability requirement is mapped to a set of replicas. Topology information containing the input/output ports of the devices is used to create multicast groups. These groups ensure the INC forwards packets to all replicas, thereby synchronizing the state of the replicas.
    \item \textbf{Defining consistency.} The consistency model to be used is configured in the servers, establishing how they should replay packets (whenever necessary for recovery). In addition, the configuration of merge functions is created by mapping commands that solve conflicts during the recovery.
    \item \textbf{Recovery.} Finally, \sysname{} creates rules to determine the need for retransmissions in case of a failure. These rules comprise a list of servers and their corresponding IPs. This list enables orchestrating the recovery by triggering route and interface changes after a failure.
\end{itemize}

%% file: evaluation.tex
\section{Evaluation}


In this section, we present experimental results to show that: (i) the refinement components of \sysname{} effectively provide fault tolerance; (ii) the system provides abstractions for reducing the overhead of recovery scenarios; (iii) the system scales for increasingly amounts of intents. 



\subsection{Experimental settings}

\textbf{Implementation.} The \sysname{} coordinator is implemented as a multithread application ($\sim$250 LoC), capable of \textit{sniffing} the network to collect devices' status information, and computing the necessary information to maintain consistency using \texttt{Scapy}. The compiler is implemented using PLY (Python Lex-Yacc), to build the \sysname{} language ($\sim$120 LoC), while also using native Linux commands to manage the repository of functionalities and instantiate experiments automatically. We built the switch building blocks using P4-16 both for V1Model ($\sim$350 LoC), and also a proof-of-concept for the TNA model ($\sim$610 LoC). We employ a specific multicast for replication, combining cloning and recirculation to buffer packets and recovery from packet loss.  \looseness=-1 

\textbf{Testbed Setup.} Our prototype for the V1Model is evaluated in a Linux virtual machine with an Intel® i5-10210U CPU @ 1.60GHz using 2 dedicated cores, 2 GB of memory, and Ubuntu 20.04 LTS. To evaluate the functionality of the system we use BMv2, a behaviour model for P4 programs. The network is emulated using mininet. The topology includes 45 hosts connected to 2 switches, one acting as a replica and the other as the main. 

We evaluate our TNA PoC of \sysname{} in a Tofino testbed. The experiments were conducted in a setup with two servers connected to two Wedge 100BF-32X 32-port programmable switches with a 3.2 Tbps Tofino ASIC. Each server is an Intel(R) Xeon(R) Silver 4210R CPU @ 2.4 GHz, with ten cores and 32 GB memory. Each server has a network interface card with two interfaces (one per switch).


\textbf{Methodology and metrics.} We run experiments to check the \textit{feasibility} of achieving fault tolerance with \sysname{}, and measure the number of requests processed per second (RPS). To understand the scalability of multiple recovery strategies, we investigate the \textit{latency} to recover from a failure and the number of packet retransmissions using different intent configurations. These measurements enable us to understand the trade-offs of different fault tolerance techniques for INC. Finally we measure the system overhead in terms of rules and resources used in the switches. \looseness=-1

\subsection{A running example}
\label{sec:runningexample}

To understand the end-to-end intent specification and refinement process, we provide a use case in our emulated setup. In this use case, we examine the intent specified in Section \ref{sec:intentspec}, Listing \ref{lst:simulation}, discuss the refinement process, and check the effectiveness of fault tolerance. Our use case deploys NetGVT \cite{parizotto2022netgvt}, an INC capable of synchronizing logical clocks from distributed systems running on servers. The nodes in the system exchange event messages with timestamps, which are intercepted and synchronized by the INC. Next, we demonstrate the step-by-step process of instrumenting this INC using \sysname{}. 

\textbf{Refinement.} Listing \ref{lst:label} presents a fragment of configurations and commands created by the \sysname{} refinement process.  The `\texttt{syncnIntent}' (Listing \ref{lst:simulation}) is refined into rules in JSON that instantiate two replicas and create a multicast group. A mirroring port is used by \sysname{} switches to clone packets and keep a copy internally in the switch. The refinement also translates the strong consistency notion to CLI commands for the switches by writing the \texttt{consistency_model} register value to 1 (corresponding to the strong consistency behavior). 

\begin{lstlisting}[aboveskip=0.2 \baselineskip, belowskip=1 \baselineskip, float=htpb, caption={Fragment of commands and configurations created.},captionpos=b, label={lst:label}]

//multicast rules created for replication
"multicast_group_entries" : [{"multicast_group_id" : 1, "replicas" : [{"egress_port" : 1, "instance" : 1}] 

//clone port for buffering
mirroring_add 500 3

//Writing specific consistency model
register_write consistency_model 0 1
\end{lstlisting}

\textbf{Fault tolerance analysis.} To analyze the functionality of \sysname{} fault tolerance, we deployed the intent and injected a failure in the switch running the INC. We then analyze the number of requests sent and acknowledged by each one of the servers.\looseness=-1

Figure \ref{fig:emulation} presents the number of requests per second processed by each server. After injecting a failure at the switch, the controller identifies the crash after a timeout. The failure leads all servers to stop transmitting packets. After the coordinator collects the devices' status to achieve consistency, the servers are notified about the failure and start the recovery by switching their communication to a different replica ($\sim16$s). Next, each server replays packets (that were lost during the switch failure) to the new main replica. After the replica finishes processing all the packets, the application returns to regular operation ($\sim18$s). \looseness=-1

\subsection{Analysing recovery configurations}

To better understand the scalability of \sysname{}, we run experiments in our emulated setup varying the number of servers and using different configuration scenarios to define the recovery strategy:

\begin{itemize}
    \item \textit{\textbf{Scenario 1}}: For the replication, the main INC periodically \textit{exchanges snapshots} with its replicas in a 4-second interval, and uses \textit{server replaying of lost packets} to achieve total order (strong consistency). 
    \item \textit{\textbf{Scenario 2}}: For the replication, the main INC \textit{sends all packets} to replicas, and uses \textit{server replaying of lost packets} to achieve total order (strong consistency). 
    \item \textit{\textbf{Scenario 3}}: For the replication, the main INC \textit{sends all packets to replicas}, but relies on a \textit{merge function and CRDTs} during recovery. The merge function solves conflicts locally at each server before the server retransmits, outputting only the last packet retransmitted before failure (strong eventual consistency). 
\end{itemize}

During the experiment, we intentionally dropped packets from the main switch to the replica, but delivered the original packet to the host destination. After a fixed interval of 4 seconds, we injected a crash in the main switch. This situation creates dependency violations that need to be corrected by the recovery procedure.   \looseness=-1

\textbf{Recovery latency.} Figure \ref{fig:Crashrecovery} presents how long the system takes to recover for each scenario. We observe that the recovery is slower as the number of servers increases. Achieving total order with eight servers requires about 7 seconds in Scenario 1 and approximately 4 seconds in Scenario 2. This latency increase is attributed to the higher number of dependencies that need correction. However, the latency does not exhibit the same growth in Scenario 3, which employs a merge function to resolve conflicts and uses Conflict-Free Replicated Data Types (CRDTs). Conflicts are resolved by consistently selecting the highest NetGVT's clock value from each server, eliminating the need to retransmit all packets. Operations to be performed on resulting packets are commutative, allowing them to be processed in any order in the replicas. This significantly reduces the number of packets requiring retransmission and avoids the need for reordering, resulting in recovery times of less than 2 seconds for any number of servers in our evaluation.

\textbf{Retransmissions and dependencies.} Figure \ref{fig:replay} presents the number of packet retransmissions due to dependency violations we observed per server in experiments with scenarios 1 and 2. We omit Scenario 3 since in this recovery strategy the dependency violations are solved by the merge function. We observe that as the number of servers increases, there is a corresponding increase in the number of retransmissions for Scenario 1. This explains the higher overhead to recover. In contrast, Scenario 2 displays a lower number of retransmissions (3 packets on average), because the failure in the INC (and the subsequent loss of packets) has a lower impact than losing entire snapshots (Scenario 1). Although reducing the number of retransmissions can improve the time to recovery compared to Scenario 1, it still requires reordering packets from multiple servers.

\subsection{Hardware micro-benchmark}

\begin{figure}[t!]
    \centering
    \includegraphics[width=0.43\textwidth]{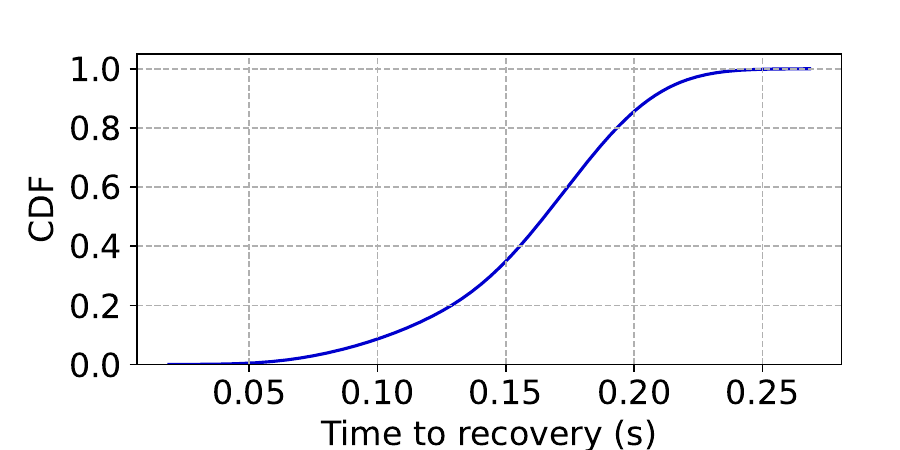}
    \caption{Failover Time.}
    \vspace{-0.3cm}
    \label{fig:enter-label}
\end{figure}

To understand the overhead of \sysname{} in real hardware, we selected the optimal configuration scenario for NetGVT (Scenario 3) and conducted experiments to measure the recovery latency. Specifically, in this experiment, we used two Tofino ASICs running \sysname{} and two servers exchanging events processed by the switches. We bring down the main switch and measure the time taken for servers to resume their operation after the recovery. The results in Figure~\ref{fig:enter-label} show that the system requires, on average, 0.16  seconds to recover from a failure. The standard deviation is 0.03 sec. These results show that \sysname{} can rapidly recover from failures while maintaining a strong notion of consistency.

\subsection{Scalability}

\begin{figure}[t!]
    \centering
    \includegraphics[width=0.43\textwidth]{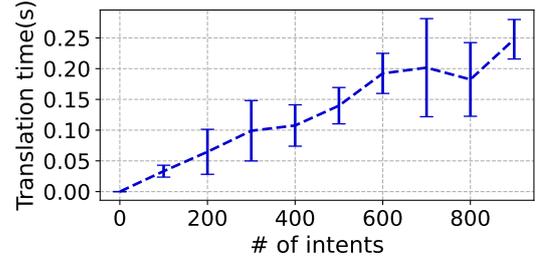}
    \caption{Time to translate intents.}
    \vspace{-0.3cm}
    \label{fig:compiletime}
\end{figure}

To show the scalability of the compiler, we generated a variable amount of intents in the \sysname{} language. In Figure \ref{fig:compiletime}, we show the time it takes to complete translation for a varying amount of intents using batches of \textit{multiple} sizes. These experiments were executed in a Linux virtual machine with an Intel® i5-10210U CPU @ 1.60GHz using 2 dedicated cores, 2 GB of memory.  \looseness=-1

We observe that the time to translate intents increases linearly with the amount of intents. Translating a single intent takes less than 0.05 seconds while translating 800 intents takes only 0.20 seconds. Overall, this result indicates that the system can translate intents rapidly.

To understand the impact on resource usage, Table \ref{tab:meas} presents the number of P4 primitives generated by the refinement process. We focus on these primitives in our evaluation because they can be generalized to multiple targets. Considering our emulated setup, which includes 45 hosts connected to 2 switches, \sysname{} has used 49 match+action entries to ensure proper communication between the coordinator and servers. The clone primitives were employed four times in the source code, enabling the creation of packet copies for acknowledgments, multicast, buffering, and retransmissions. A single recirculation primitive was used for buffering packets. These results indicate that only a small amount of P4 primitives are used by \sysname{}.


\begin{table}[t!]
\caption{Rules and primitives used by \sysname{}.}
\label{tab:meas}
\centering
\resizebox{0.27\textwidth}{!}{%
\begin{tabular}{|l|c|}
\hline
\multicolumn{1}{|c|}{\textbf{Primitives/Rules}} & \textbf{Usage} \\ \hline \hline
Match+Action Entries                            & 49             \\ 
CloneE2E                                        & 4              \\ 
Multicast Groups                                & 1              \\ 
Recirculation                                   & 1              \\ \hline
\end{tabular}%
}\end{table}

%% file: related_work.tex
\section{Related Work}

\textbf{Intents.} Various platforms have been explored for intent management, including network function virtualization \cite{scheid2017inspire}, software-defined networking \cite{jacobs-2018, 8725754, 10.1145/3281411.3281431, alalmaei2020sdn}, and industrial networks \cite{saha2018intent}. Researchers have been investigating techniques for managing Access Control Lists (ACL) \cite{tian2019safely, li2021automatic} and Quality of Service (QoS) using intents, along with exploring diverse abstractions to express intents. These abstractions include policy graphs \cite{Prakash:2015:PUG:2785956.2787506}, natural language \cite{jacobs-2018}, graphical user interfaces \cite{femminella2020simplification}, and constrained natural language grammars \cite{scheid2020controlled}. Recent works also used high-level intents to create \textit{match+action} entries for P4 programs \cite{angi2022nlp4, 8526852}. P4I/O \cite{riftadi2019p4i} facilitates P4 adoption by using a language to express policies for switches, and merges different source files using reusable templates to dynamically upgrade the switch configuration. However, the authors only demonstrated the concept for deploying heavy hitter detection. \sysname{} focuses on other domain by extending INC functionality with specific abstractions for fault tolerance.

\textbf{High-level data structures.} An orthogonal research field focuses on bringing higher-level abstractions to P4. Examples of high-level abstractions include data structure elasticity \cite{hogan2022modular}, loops \cite{alcoz2022reducing}, modularity \cite{fattaholmanan2021p4}  composability \cite{soni2020composing}, and heterogeneity \cite{gao2020lyra}. Although these efforts can simplify programming, they do not support functionalities to express intents. \looseness=-1

\textbf{INC Fault tolerance.} Fault tolerance is investigated in \cite{jin2018netchain, huang2022hypersfp}, focusing on applications like aggregation, key-value store, and network functions. Notably, RedPlane \cite{kim2021redplane} proposes techniques for making switch applications fault tolerant. RedPlane synchronizes the switch with state storage to provide strong and eventual consistency. Swish \cite{zeno2022swish} provides abstractions that can enable distributed network functions on programmable switches by replicating the state and operations between multiple devices. However, none of the previous approaches support the specification of requirements at the level of intents. 





%% file: conclusions.tex
\section{Conclusion}

In this work, we presented \sysname{}, a system to provide fault tolerance requirements expressed as intents for INCs. The system allows intents to be specified in a constrained natural language. Subsequently, a refinement mechanism instruments the INC code for ensuring fault tolerance. We have implemented a prototype of \sysname{} both in an emulated setup based on BMv2 and on a Tofino testbed, and analyzed our translation and refinement processes. As future work, we plan to address other kinds of failures, including adversarial attacks, bugs and malfunctioning. Simplifying intent management using large language models is also an interesting perspective. Furthermore, we also plan to investigate how eBPF can be used to reduce the overhead of recovery. 